\documentclass[aps,prl,twocolumn]{revtex4-1}
\usepackage{graphicx}
\usepackage{bm}
\usepackage{amsbsy}
\usepackage[usenames,dvipsnames,svgnames,table]{xcolor}
\usepackage{mathrsfs,mathcomp}
\usepackage[normalem]{ulem}
\usepackage{amsmath}
\usepackage{amsfonts}
\usepackage{amssymb}

\usepackage{amsfonts,amsmath,amssymb}
\usepackage{textcomp}
\usepackage{bm}
\usepackage{graphicx}
\usepackage{rotating}
\usepackage{color}
\usepackage{hyperref}
\usepackage{siunitx}

\newcommand{\avr}[1]{\langle #1 \rangle}

\newcommand{\beq}[1]{\begin{equation} \eqlab{#1}}
\newcommand{\eeq}{\end{equation}}
\newcommand{\bsub}{\begin{subequations}}
\newcommand{\esub}{\end{subequations}}
\def\bal#1\eal{\begin{align}#1\end{align}}
\def\bsubal#1\esubal{\bsub \begin{align}#1\end{align} \esub}

\newcommand{\eqlab}[1]{\label{eq:#1}}
\renewcommand{\eqref}[1]{Eq.~(\ref{eq:#1})}

\pdfoutput=1

\usepackage{color}

\newcommand{\ed}[0]{{\rm e}}

\begin{document}


\title{Clogging in constricted suspension flows}

\author{Alvaro Marin$^{*a}$, Henri Lhuissier$^b$, Massimiliano Rossi$^c$, Christian J. K\"ahler$^c$} 

\affiliation{$^a$Physics of Fluids, University of Twente, The Netherlands\\
$^b$ Aix Marseille Univ, CNRS, IUSTI, Marseille, France\\
$^c$Institut f\"ur Str\"omungsmechanik und Aerodynamik, Bundeswehr University Munich, Germany}

\begin{abstract}
	The flow of a \textcolor{black}{charged-stabilized} suspension through a single constricted channel is studied experimentally by tracking the particles individually. Surprisingly, the behavior is found to be qualitatively similar to that of inertial dry granular systems: For small values of the neck-to-particle size ratio ($D/d<3$), clogs form randomly as arches of particle span the constriction. The statistics of the clogging events are Poissonian as reported for granular systems and agree, for moderate particle volume fraction ($\phi\approx20\%$), with a simple stochastic model for the number of particles at the neck. For larger neck sizes ($D/d>3$), even at the largest $\phi$ ($\approx60\%$) achievable in the experiments, an uninterrupted particle flow is observed, which resembles that of an hourglass. This particularly small value of $D/d$ ($\simeq3$) at the transition to a practically uninterrupted flow is attributed to the low effective friction between the particles, achieved by the particle's functionalization and lubrication.
\end{abstract}

\maketitle


	Experience shows that rigid particles forced though a narrow constriction may either flow steadily,  intermittently, or do not flow at all, when the particles form a clog obstructing the route \cite{Zuriguel14}. Such flowing modes are encountered in different systems involving particle retention or discharge, such as the hopper of an emptying granular silo \cite{Beverloo61,Kulkarni10}, the neck of an hourglass \cite{Wu93}, micro-fluidic and filtration circuits, where a loaded liquid enters a device or permeates a membrane \cite{Sharp05,Lin09,Bacchin11}, but also crowds of people evacuating a room or a hungry herd entering a room through a door \cite{Helbing00,Zuriguel14}. However, besides their apparently similar behaviors, these systems conceal fundamentally different intermittency and clogging mechanisms. Hourglasses and silos contain dry granular media flowing collectively from a sedimented state to a gravity-driven flow limited by particles interactions \cite{Beverloo61}. They clog when a stable, static arch of particles spanning the constriction forms \cite{To01}. The clogging probability is essentially prescribed by the neck-to-particle size ratio, $D/d$, and the particle shape \cite{Zuriguel05}, independently from the flow rate \cite{Dorbolo10}, and almost independently from the constriction angle \cite{To01} and other particle properties \cite{Zuriguel05}. For ratios $D/d$ above typically 5 to 8, both 2D and 3D flows are virtualy uninterrupted \cite{To01,Zuriguel05}, although the critical status of this empirical assessment is still under debate \cite{Zuriguel05,To05,Masuda14,Thomas15}.
\begin{figure}
	\includegraphics[width=0.42\textwidth]{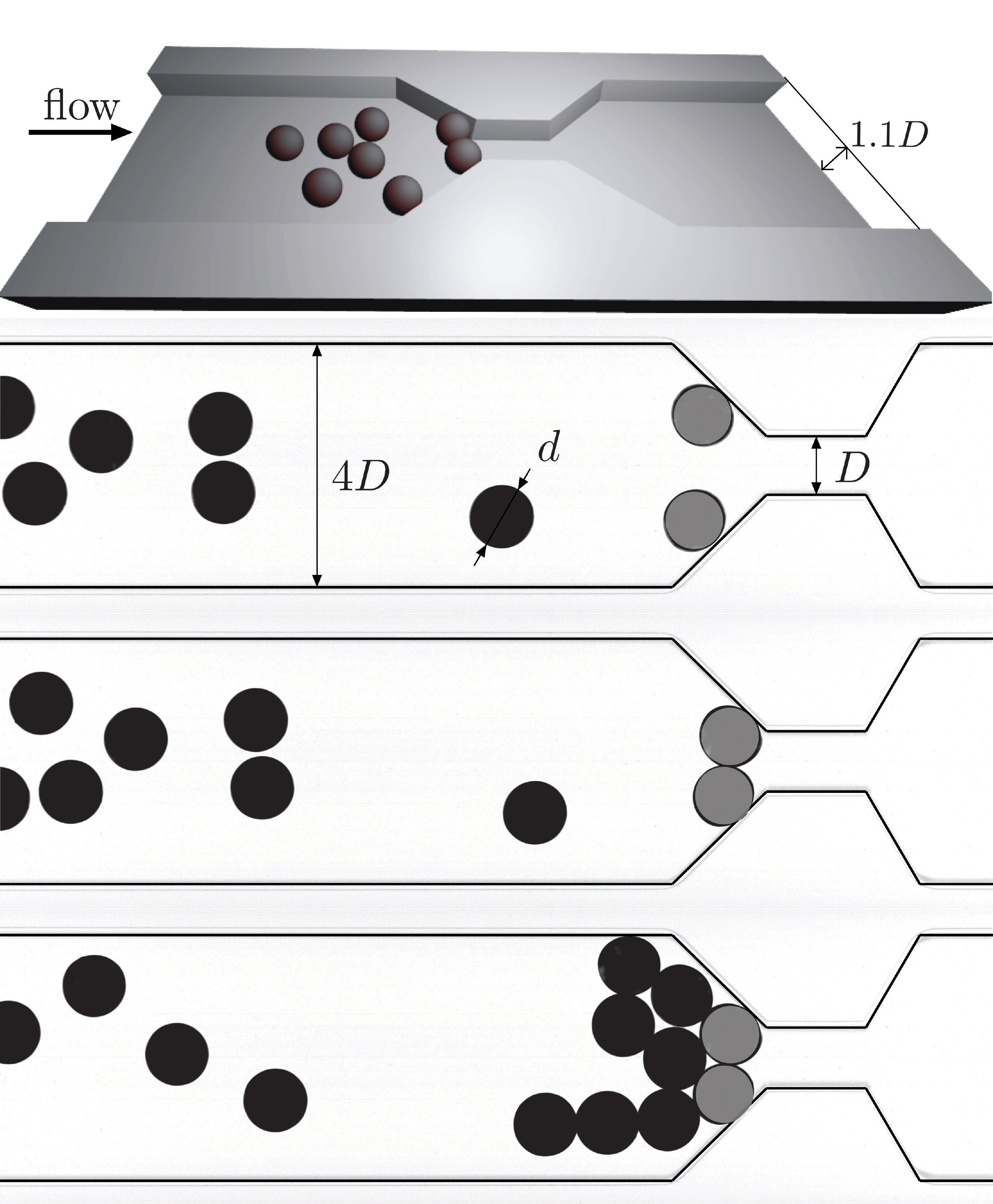}
	\caption{Top: Schematics of the system; a laterally-constricted channel with uniform thickness. A suspension of particles with a diameter $d$ smaller than the neck width $D$ and nominal particle volume fraction $\phi$ is forced through the constriction until a clog forms. Bottom: Image sequence showing the formation of a clog for $D/d = 1.02$ when a sufficient number of particles (here two) reach the constriction at the same time.}\label{fig:setup} 
\end{figure}

\begin{figure*}[t!]
	\centering
	\includegraphics[width=.98\textwidth]{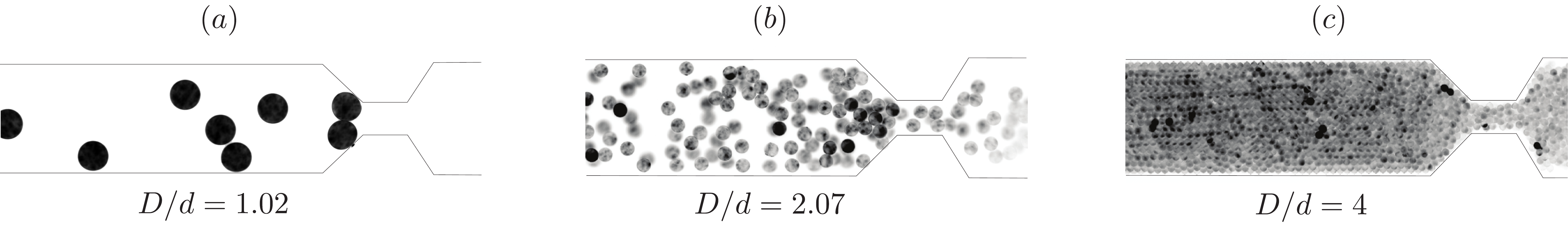}
	\caption{Snapshots of the flow at the constriction. Images (a-b) show the configuration immediately after a clog has formed for $D/d=1.02$ and $D/d=2.07$. Panel (c) shows the continuous and uninterrupted flow observed for $D/d=4$: even at the largest particle volume fraction ($\phi \approx 60\%$) no clog is observed (i.e. after {$\sim 10^8$} particles through the constriction).} \label{fig:exps}
\end{figure*}
By contrast, suspensions accommodate particle dilution and, for colloidal systems, often experience other clogging mechanisms involving single-particle effects mediated by wall/particle adhesion or particle aggregation \cite{Wyss06,Bacchin11,Agbangla14}. At large concentrations, shear-induced particle migration becomes increasingly important \cite{Karnis66,Moraczewski06}, and may, for instance, impede blood circulation in the smallest venules \cite{Zhou05}. However, suspensions have also been reported to sustain clogging by arching, qualitatively similar to silos clogging \cite{guariguata2012jamming,lafond2013orifice,Durian2017}, even at low volume fractions \cite{Ramachandran99} and in the absence of adhesion/aggregation \cite{Roussel07}. Although it is not always considered, this arching mechanism is crucial since it is expected to control clogging for suspensions of non-attractive particles. Moreover, the apparent similarity with silos is particularly surprising given the strong differences with a classical granular system: (1) The particle flow is driven by the viscous drag of the suspending liquid. (2) The particle concentration is variable. (3) Distant particle interactions, whether hydrodynamic or not, are present.

	To determine the prevalence of clogging by arching in suspensions and the possible similarities with its dry counterparts we investigate the flow of a dilute suspension of charge-stabilized non-Brownian particles through a single micro-fluidic constriction with a controlled geometry. Using high-resolution and high-speed optical video microscopy, we track individual particles as they travel throught the channel. This permits direct visualization of the clogging mechanism, as well as obtaining the statistics of the clogging events by monitoring the precise number of particles that pass through the constriction before a clog forms.


	The fluidic system consists of a straight channel of borosilicate glass (isotropic wet etching-fabricated by Micronit micro-fluidics), which is locally constricted in the middle (see Fig. \ref{fig:setup}). The channel has an uniform thickness of 110\:\textmu m. The constriction is achieved by a linear narrowing of the channel, with a half-angle of 45$^\circ$, from the nominal width of 400\:\textmu m down to 100\:\textmu m at the neck. This forms a two-dimensional nozzle converging towards the almost square cross-section of the neck ($110\times100$\:\textmu m$^2$ $-$ note that the flow itself is not two-dimensional due to the boundary conditions). The suspensions consist of monodisperse polystyrene particles stabilized with negatively-charged sulfate groups (Microparticles GmbH) in a density-matched $21.5\%$wt aqueous solution of glycerine (with a viscosity of 1.8\,mPa\,s). Different suspensions with particle diameters $d = 98.5$, 59, 48, 41, 33 or 25\:\textmu m ($\pm 2\%$) were used. Adopting the neck width, $D = 100$\:\textmu m, as the characteristic length scale, these correspond to neck-to-particle ratios $D/d=1.02$, 1.7, 2.07, 2.43, 3.03 and 4, respectively. The charged sulfate groups confer them a small negative surface potential (of order -50\:mV) but sufficient to prevent both their agglomeration and their adhesion to the channel walls (borosilicate glass). The suspensions are prepared with a particle volume fraction of about $2\%$, pre-concentrated up to approximately 20\% and inserted in the device. However, the homogeneity of the volume fraction in the channel cannot be controlled precisely, mainly due to the successive clogging and flow reversals needed to acquire the statistics (see below). We therefore measure the particle volume fraction $\phi$ in situ, close to the constriction, by directly imaging and counting the particles. This yields $\phi = 20\pm10\%$ for all $D/d$ smaller than 3, where the $\pm$ refers to slow variations in time (the time-averaged particle volume fraction was found to be uniform in the channel). For the largest particle-to-neck ratios ($D/d>3$), $\phi$ was eventually increased up to $\approx60\%$ in order to promote clogging events (see below). 
	The suspensions are driven through the constriction by a pressure-based flow controller (Fluigent) at a Reynolds number smaller than $10^{-2}$ (corresponding to particle velocities $\gtrsim1\,$mm/s). Identical results were obtained with a volume-driven flow using a syringe pump. Note that most of the data has been obtained with dilute suspensions ($\phi\approx 20\%$), for which the particle flow rate is set only by the liquid flow rate and the particle volume fraction. 
\begin{figure*}[]
	\includegraphics[width=0.4\textwidth]{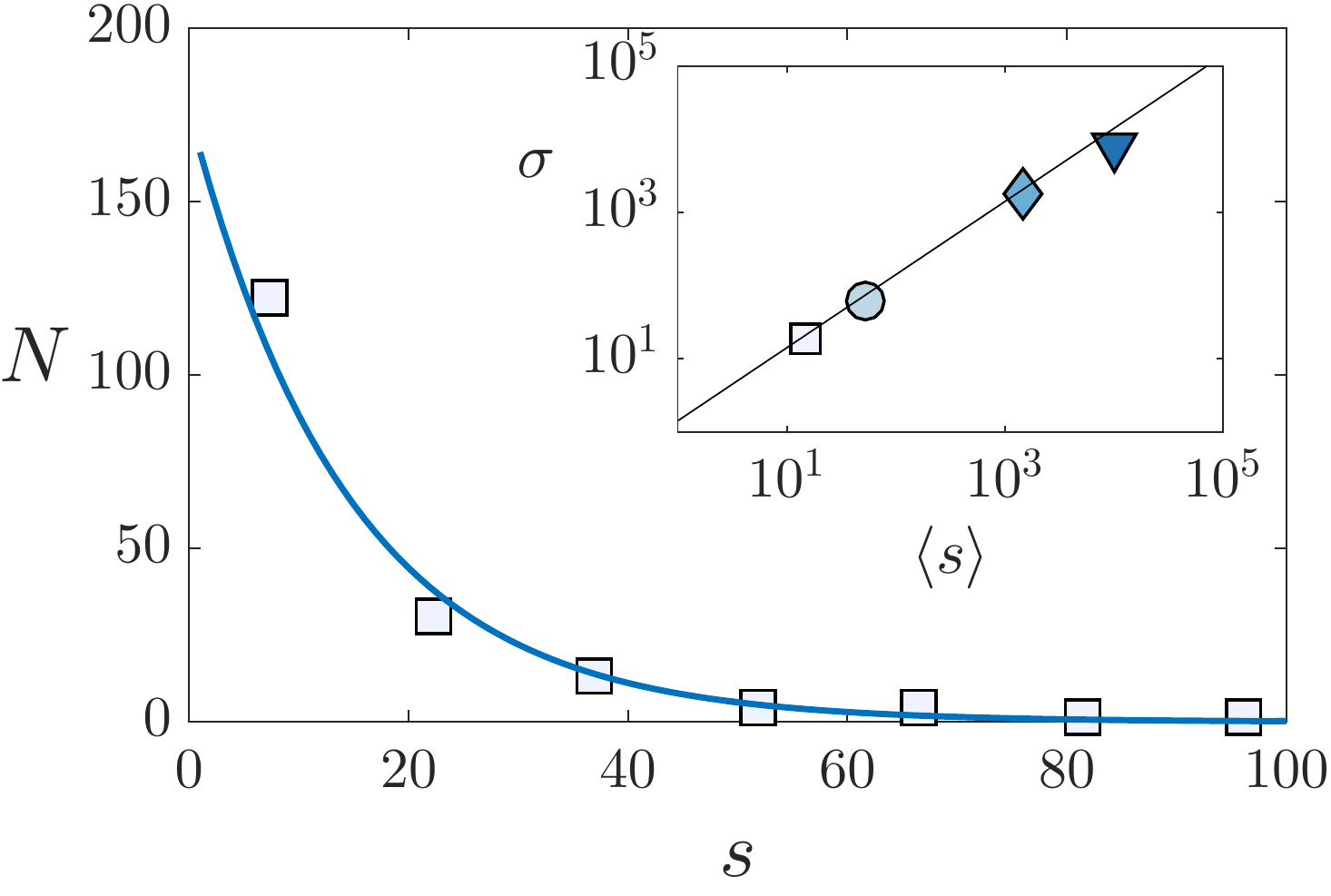}\hspace{1cm}
	\includegraphics[width=0.4\textwidth]{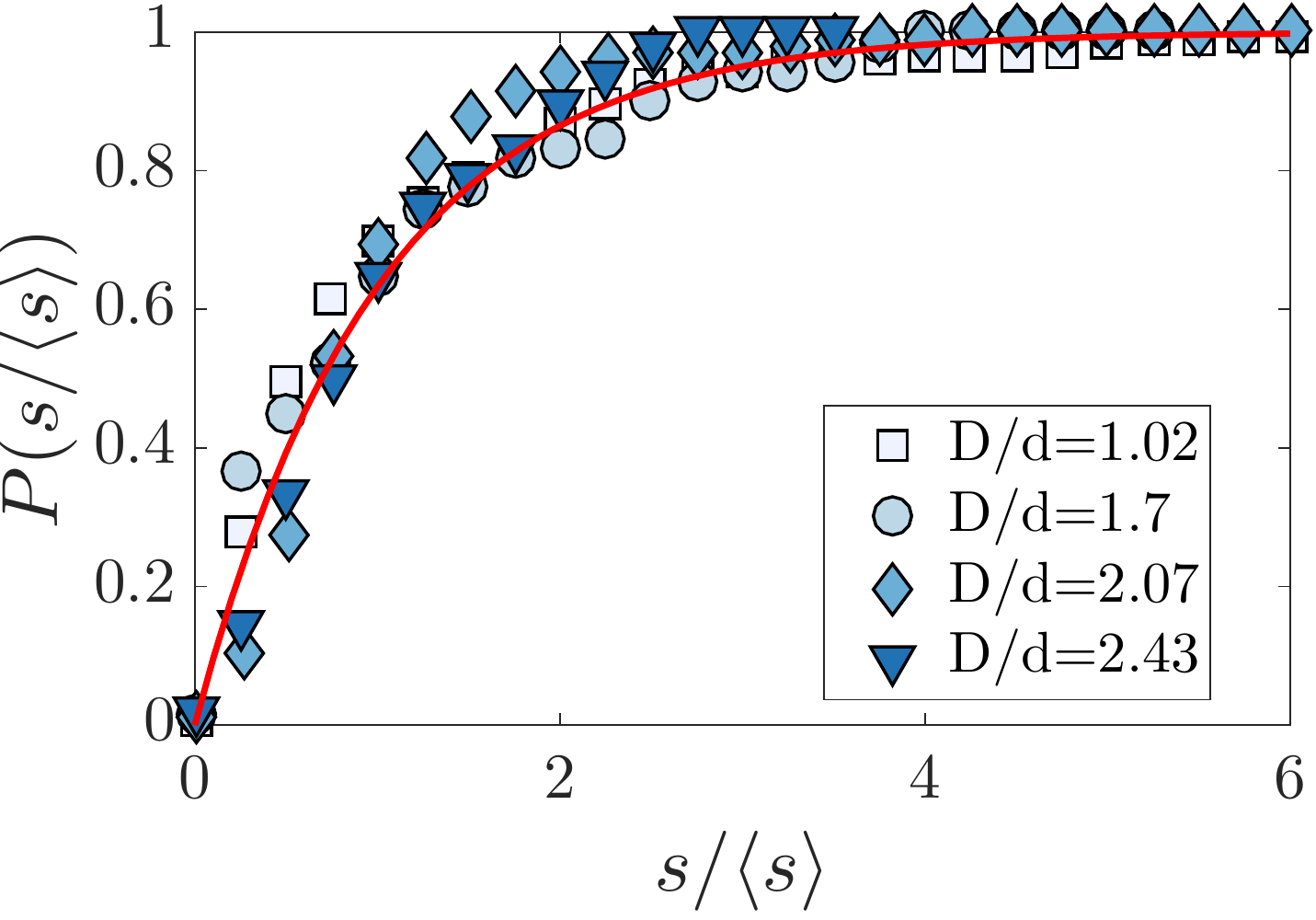}
	\caption{(Left) Histogram of the number of escaped particles $s$ for $D/d=1.02$. The number of events $N$ decays with increasing $s$. The line represents the exponential distribution $\sum_{s'=0}^\infty N(s')\avr{s}^{-1}{\rm e}^{-s/\avr{s}}$, having the same mean value $\avr{s} = 14.5$ as the data. Insert: Standard deviation of the number of escapees $\sqrt{\avr{s^2}}$ compared to the value $\sqrt{2}\avr{s}$ (solid line) expected for an exponential distribution for $D/d=$ 1.02 ($\blacksquare$), 1.7 (\textcolor[rgb]{0.3,0.4,0.4}{\large$\bullet$}), 2.07 (\textcolor[rgb]{0.65,0.7,0.7}{$\blacklozenge$}), 2.43 ($\triangledown$).\label{fig:histogram} (Right) Cumulative clogging probability of the relative number of escapees $s/\avr{s}$ for neck-to-particle size ratios covering a $10^3$-fold range of mean number of escapees $\avr{s}$ (see Fig. \ref{fig:model}). The solid line is the curve $P=1-e^{-s/\avr{s}}$ expected for an exponential distribution. \label{fig:P}}
\end{figure*}

	The experiment starts when the flow is switched on and particles are dragged by the flow towards the constriction. We count the number of particles passing through the constriction until the channel clogs, which we note as $s$ and refer as \emph{number of escapees}. Due to the low adherence of the particles to the walls and to each other, the arch is fully disassembled by reversing the flow for a short time. The forward flow is then restored and a new independent clogging event is observed (the independency was verified from the absence of correlation between successive clogging events, see supplementary materials). For each of them, approximately 100 clogging events were analyzed. A typical histogram of $s$ is shown in Fig. \ref{fig:histogram}. The number $N$ of events observed for a given number of escaped particles clearly decreases with increasing $s$ and the experimental probability $N(s)/\sum_{s'=0}^\infty N(s')$ is well fitted by an exponential distribution,
\begin{equation}
	p(s) =  \frac{1}{\avr{s}}\rm{e}^{-s/\avr{s}},
	\label{eq:p}
\end{equation}
	which suggests that the clogging events can be described as a random Poisson process, as already observed for dry granular systems \cite{To05,Zuriguel05}. This assessment is also supported by analyzing the standard deviation of the number of escaped particles, $\sqrt{\avr{s^2}}$, shown in the insert of Fig. \ref{fig:histogram}. which is found to be close to the value $\sqrt{2}\avr{s}$ expected for an exponential distribution. Such behavior has been found for all experiments with $D/d<3$, in which clogging events could be observed.
	
	The data for $D/d<3$ are summarized in Fig. \ref{fig:P} in terms of the cumulative probability $P(s/\avr{s}) = \sum_{s'=0}^s p(s')$, i.e., the probability of having less than $s$ particles passing through the constriction. As shown in Fig. \ref{fig:P}, the probabilities, normalized by the average number of escapees $\avr{s}$, collapse for all the data sets. This suggests that particles, or small groups of particles, pass through the constriction independently from each other and have all the same probability of clogging the constriction, regardless of the history of clogging events. Note that the physical reason underpinning this independency is not the same as for dry granular flows. Whereas for granular media it results from the saturation of the lithostatic pressure at the scale of the silo aperture \cite{Wu93}, it is, presently, a mere consequence of the moderate particle volume fraction of the suspension. Despite this difference, the statistical analogy is strong and the clogging statistics are entirely given by $\avr{s}$.
	
\begin{figure}[t]
	\includegraphics[width=.4\textwidth]{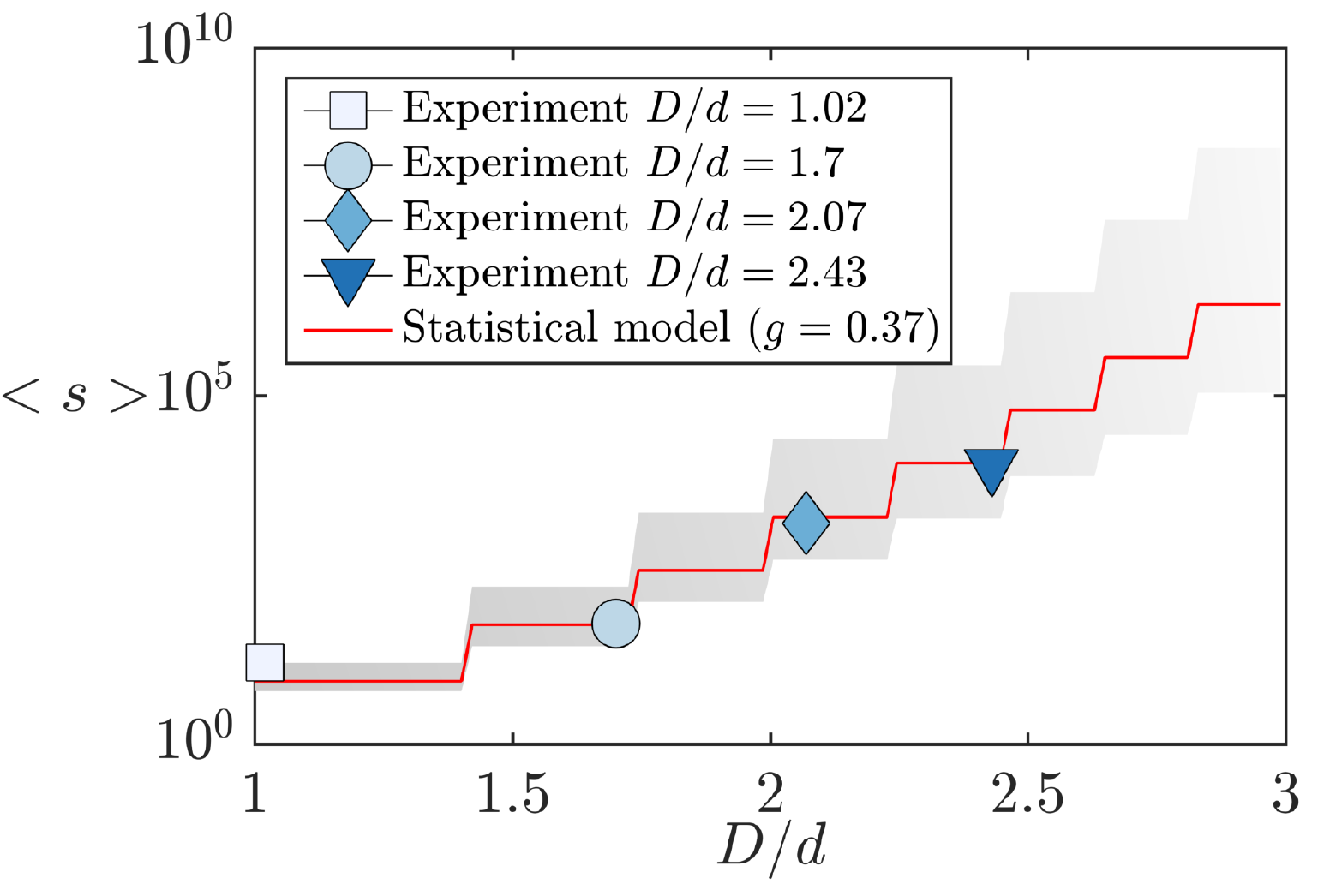}
	\caption{Average number of escapees $\avr{s}$ observed experimentally and predicted by the stochastic model, \eqref{p2} with {$g(\mu)=0.37$}. The solid line refers to the prediction for the mean particle volume fraction $\phi = 20\%$. The gray band represents deviations from the predicted $\avr{s}$ value due to unintended changes in volume fraction $10\%<\phi<30\%$.\label{fig:model}}
\end{figure}

	We therefore turn to the mean number of escapees $\avr{s}$. For the typical flow rate considered here ($\sim 1\,$mL/hr), $\avr{s}$ is found to be practically independent on the flow rate ({see supplementary materials}), but strongly dependent on $D/d$. Not surprisingly, $\avr{s}$ increases, i.e. clogging events become less likely, as the particles become smaller. For $D/d=1.02$ (Fig. \ref{fig:exps}a), a clogging event typically occurs after only $\avr{s}\approx 15$ particles have passed through the aperture. By doubling $D/d$ to $2.07$ (Fig. \ref{fig:exps}b), the average number of escapees $\avr{s}$ rises to as much as $\approx1500$. As Fig. \ref{fig:model} shows, this steep increase with $D/d$ perseveres for larger values of the neck-to-particle size ratio. Actually, for $D/d=4$ the probability of clogging is so small that no permanent clogs could be observed $-$ hence no statistics of $s$ could be obtained. More surprisingly, even after increasing $\phi$ up to the largest particle volume fraction $\phi\approx60\%$ achievable in our system, the flow could be maintained over several hours at a volume flow rate of $\sim1$\:mL/hr without any interruption. That is to say, no persistent clog had formed after typically {$10^8$} particles had passed through the constriction (implying {$\avr{s} \gtrsim 10^8$}). For most practical purposes such flow can probably be considered as uninterrupted, which should be all the more true for lower $\phi$ since $\avr{s}$ is expected to be even larger. It is however not possible to conclude on whether the clogging probability has identically vanished or if it has only dramatically decreased to a value smaller than $10^{-8}$ despite $\phi\approx60\%$. Also, we must precise that although no permanent clog is observed for $D/d>3$, different flows develop at large $\phi$. For $D/d=3.03$ the flow is intermittent, i.e. particle flow occurs in erratic bursts, separated by short periods of arrest (of a few seconds). During each particle burst, approximately $10^4-10^5$ particles can typically escape. Whereas for $D/d>4$ the flow is continuous (Fig. \ref{fig:exps}c), as already observed with dry grains (hourglass) \cite{Wu93} and colloidal suspensions \cite{Genovese11}.

	In order to explain the clogging behavior of the constriction, we now proceed to build a simple statistical model describing the average number of escapees $\avr{s}$ for low particle volume fractions. We base our model on two observations. First, in the present case of non-adhesive particles, the particles do not aggregate and remain close to randomly dispersed in the suspending liquid. Second, arches do not build up progressively from the contribution of successive particles \cite{Wyss06}, but form suddenly. As shown in Fig. \ref{fig:setup}, a stable arch is formed when a sufficient number of particles $n$ reach the constriction within a sufficiently short time interval, that is to say, when $n$ particles are contained at the same time in a given critical volume $\Omega$ prescribed by the geometry of the constriction and the particles properties. Now, the probability that among particles with a nominal number density $c\equiv6\phi/\pi d^3$ and animated by random and independent motions, exactly $n$ particles lie at the same time in a given arbitrary volume $\Omega$ is
\begin{eqnarray}
	q = \frac{(\Omega c)^{n-1}}{(n-1)!}\,\ed^{-\Omega c}.
	\label{eq:q}
\end{eqnarray}
	This probability strongly decreases with increasing values of $n$. The probability of forming a clog is therefore expected to be dominated by the minimal number of particles required to form an arch, i.e, by 
\begin{eqnarray}
	n \simeq \frac{D^2}{d^2} + 1,
	\label{eq:n}
\end{eqnarray}
	assuming the minimal stable arch typically spans the neck cross-section. The critical volume $\Omega$ in which these particles must lie to form a clog on their arrival at the constriction can be thought of as the typical volume span by the particles center in all possible conformations corresponding to a stable arch (see \cite{To01}). $\Omega$ is expected to depend upon $n$, as well as upon the particle interactions and the angle of the constriction. We assume that it is essentially set by the volume of the $n$ particles modulated by an effective friction coefficient $\mu$ through a geometrical pre-factor $g(\mu)$ according to
\begin{eqnarray}
	\Omega = g(\mu) \times n \frac{\pi d^3}{6}.
	\label{eq:Om}
\end{eqnarray}

	In the limit $s\gg n$, the probability $p(s)$ that a clog forms after the escape of exactly $s$ particles simply verifies $\partial_s (p/q) = -p$. Making use of Eq. (\ref{eq:p}) and (\ref{eq:Om}) and considering a steady nominal volume fraction ($\phi = \rm{cte}$), this yields the following distribution and average number of escapees:
\begin{eqnarray}
	p(s) &=& q\, \ed^{-qs},\\
	\langle s\rangle &=& q^{-1} = \frac{(n-1)!}{(g \phi n)^{n-1}}\,\ed^{g \phi n}.
	\label{eq:p2}
\end{eqnarray}
Together with \eqref{n}, this makes the neck-to-particle ratio $D/d$ explicit.

	For a given $D/d$, \eqref{p2} recovers the experimental exponential distribution of $s$ given in \eqref{p}. It also predicts a steep increase in the average number of escapees as a function of the neck-to-particle ratio, since $\avr{s} \sim (D/d)(g \phi\ed^{1-g\phi})^{-(D/d)^2}$. \eqref{p2} is compared to the experimental statistics in Fig. \ref{fig:model}, for all values of $D/d$ for which clogs have been observed ($D/d<3$). Considering the nominal particle volume fraction $\phi = 0.2 \pm 0.1$, a single parameter remains, namely, the geometrical function $g$ appearing in $\Omega$. Taking $g(\mu)=0.37$, and besides its crude assumptions, the model predicts surprisingly well the experimental results over the whole range of neck-to-particle size ratios.




	To conclude, we have shown that non-adhesive colloidal particles in suspension flow across constrictions in a strikingly similar way as  dry non-cohesive granular systems do. 
	At low Reynolds number and in the absence of particle aggregation or particle/wall adhesion, the neck-to-particle size ratio, $D/d$, determines the flow regime (clog, intermintent or continuous), independently from the flow rate. At low $D/d$, the flow is interrupted by the formation of stable particle arches spanning the constriction. The clogging events follow Poisson statistics which can be described, in the diluted regime, with a simple stochastic model for the number of particles forming the arch. Interestingly, this stochasticity has a different origin as for dry granular flows. Whereas for the latter the loss of memory is a consequence of the multiple particle collisions, in the present dilute case it simply stems from the random arrival of the particles at the neck. At larger $D/d>3$, the flow becomes uninterrupted over the longest experimental duration achievable ({$\sim 10^8$} particles), even upon dramatically increasing the particle volume fraction to $\phi\approx60\%$. This apparently continuous flow regime appears for lower values of $D/d$ than for dry granular systems \cite{To01,Zuriguel05}. This is consistent with the decrease in the critical $D/d$ with decreasing particle/particle friction coefficient reported for dry systems \cite{To01,To05}, if one assumes that the lubrication forces and the repulsive dielectric forces maintain a liquid layer between the particles, which results in a low effective friction \cite{Clavaud17}. This suggests that, even in the absence of adhesion, the flow of particles through confined geometries can be significantly facilitated by engineering short-range repulsive forces (of electrostatic or polymer/surfactant-induced nature) and smoother surfaces. Such a feature could be exploited to enhance the functional range of fluidic devices, filters and membranes intended for suspensions.

\paragraph{Acknowledgments.-}
	The authors acknowledge the technical support of Andreas Volk in the experiments. The authors acknowledge financial support by the Deutsche Forschungsgemeinschaft (DFG) grant KA1808/12.

\bibliographystyle{unsrt} 

\end{document}